\documentclass[10pt,a4paper]{article} 

\usepackage{anysize}
\usepackage[format = hang]{caption}
\usepackage{amsmath,amsthm,verbatim,amssymb,amsfonts,amscd, graphicx}
\usepackage{graphics,natbib}
\usepackage{titlesec,footmisc}
\usepackage{hyperref} 

\usepackage{listings}
\usepackage{booktabs} 

\theoremstyle{plain}

\theoremstyle{definition}

\newcommand{\corr}{\text{corr}}

\def\E{{\rm E}}
\def\Var{{\rm Var}}

\def\tr{{\rm t}}

\def\hatt{\widehat}
\def\arr{\rightarrow}
\def\N{{\rm N}}
\def\half{\hbox{$1\over 2$}}

\def\emp{{\rm emp}}
\def\Gam{{\rm Gamma}}

\def\se{{\rm se}}

\def\beq{\begin{eqnarray}}
\def\eeq{\end{eqnarray}}

\def\beqn{\begin{eqnarray*}}  
\def\eeqn{\end{eqnarray*}}

\def\E{{\rm E}}
\def\Var{{\rm Var}}

\def\N{{\rm N}}

\def\Pr{P}

\def\quadandquad{\quad {\rm and} \quad}
\def\arr{\rightarrow}
\def\hatt{\widehat}

\def\sumin{\sum_{i=1}^n}

\def\maxin{\max_{i\le n}}

\def\half{\hbox{$1\over2$}}

\def\obs{{\rm obs}}

\def\tr{{\rm t}}

\def\sd{{\rm sd}}

\titleformat{\section}{\normalfont\large\sc\centering}{\thesection}{1em}{}
\titleformat{\subsection}[runin]{\normalfont\large\bfseries}{\thesubsection}{1em}{}
\numberwithin{equation}{section} 
\renewenvironment{abstract}
               {\list{}{\rightmargin\leftmargin}%
                \item[\text{\hspace{10mm}\sc Abstract.}]\relax}
               {\endlist}

\begin{document}

\def\heute{January 7, 2026}

\begingroup
\begin{centering} 

  \Large{\bf Anyone for chess? \\ Analysing chess ratings above
  high thresholds }\\[0.8em]
\large{\bf Nils Lid Hjort} \\[0.3em] 
\small {\sc Department of Mathematics, University of Oslo} \\[0.3em]
\small {\sc {\heute}}\par
\end{centering}
\endgroup


\begin{abstract}
\small{Suppose some cleverness score parameter is sufficiently
    interesting to be defined and then measured, perhaps for
    different strata of specialists or for the broader population.
    Such phenomena could have Gaussian distributions,
    when it comes to all players in a stratum, but when interest
    focuses on the very tails, for the top few percent,
    those above certain high thresholds, 
    different models are called for, along with the need
    to analyse such based on the listed top scores only.
    In this note I develop such models and tools,
    and apply them to the top-100 and above 2100 points
    lists for regular chess ratings, for the currently active
    14671 men and 753 women,
    as given by the FIDE, January 2026.
    It is argued that even when two or more distributions have
    close to identical expected values, or medians,
    even smaller differences in variance may explain gaps
    for the few very best ones.
    
}
\end{abstract}


\section{Introduction: the sizes of maxima}
\label{section:maxima}

\begin{figure}[h]
\centering
\includegraphics[scale=0.17]{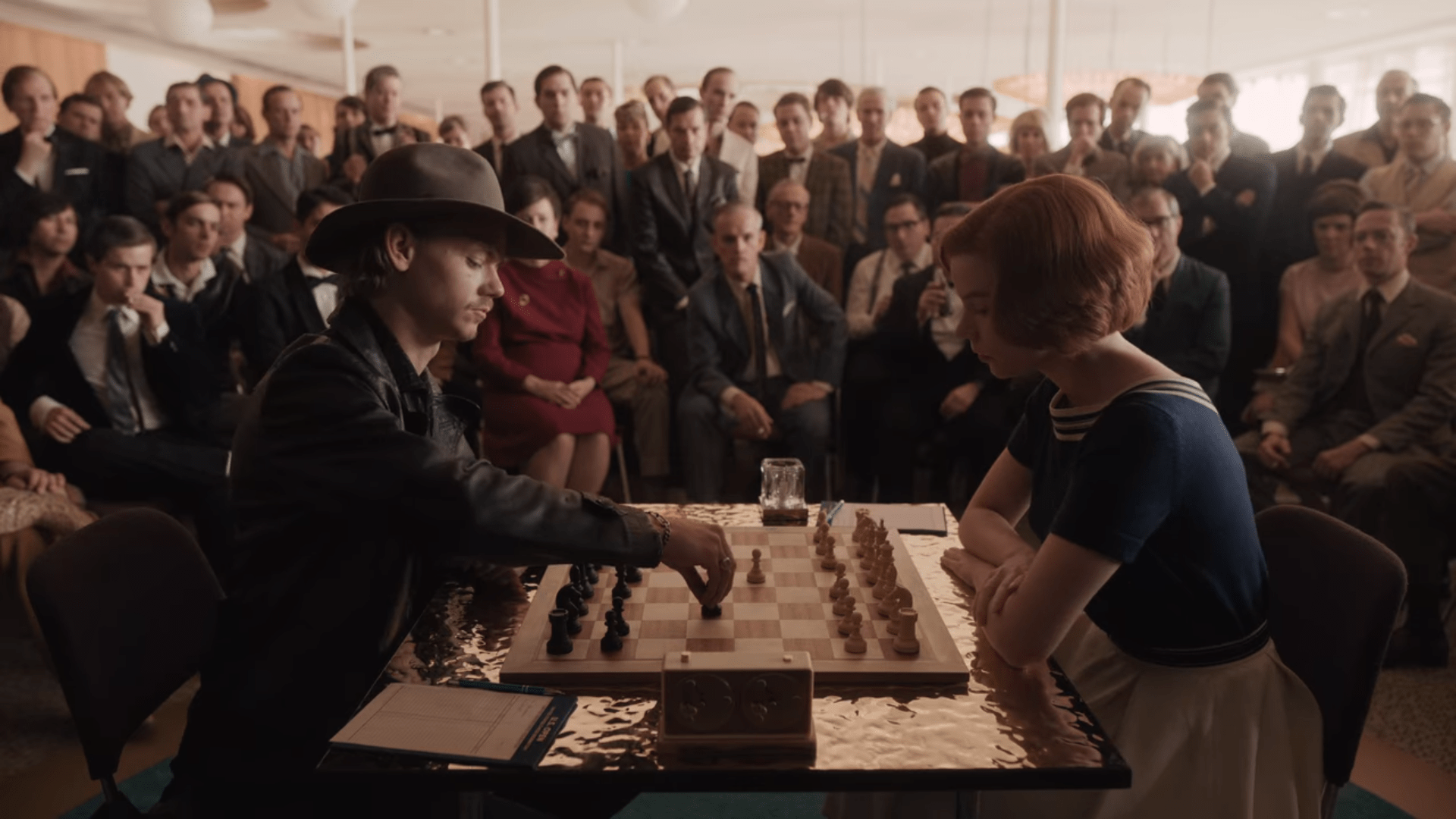} 
\caption{
  Queen's Gambit.}
\label{figure:queen}
\end{figure}

\noindent 
The beautiful game of chess is for nearly everyone:
just grab the pieces and your nearby board,
on your laptop or on your phone, or perhaps your analogue bishops
and rooks and knights and pawns, and play on.
It takes beautiful minds to play it really
wondrously, however.
``Not In 10,000 Lifetimes Will We See Chess Like This Again'',
proclaims {\it Epic Chess}, among the many chess experts
on the net, with hundreds of thousands of followers,
analysing in detail Magnus Carlsen's complicated march
towards winning the World Rapid and World Blitz, in Doha,
Christmas week 2025. So statisticians should be interested too. 

In this note I study the chess ratings,
as per the El\"o scale used by the International Chess
Federation FIDE in clubs and tournaments all over the world,
not on the population scale but for the very clever ones,
those with ratings above a high threshold, like $r_0=2100$.
Before I come to chess, however,
with models and analysis methods developed in
Section \ref{section:paramodel}, it is useful to build
some intuition around extreme scores, under natural and simple
sets of assumptions. The aim is partly to identify
what the driving factors are, for extreme scores
inside different strata. 


Consider therefore $M_n$, the maximum of $X_1,\ldots,X_n$,
an i.i.d.~sample from the $\N(\xi,\sigma^2)$. We may think
in terms of a stratum of people and their heights,
with $M_n$ the height of the tallest. Then, with
$c_n\doteq d_n$ indicating $c_n/d_n\arr1$, 
\beqn
\E\,M_n\doteq\xi+\sigma(2\log n)^{1/2}, 
\eeqn 
which captures the essential ways in which the tallest can
be very tall (or perhaps not so very tall, after all):
(i) $\xi$ should be big;
(ii) higher participation, i.e.~a bigger $n$, matters,
but only on the very slowly growing $(\log n)^{1/2}$ scale; and
(iii) it is also helpful with a not small $\sigma$.
There is a similar approximation also in the equicorrelation case,
where individuals in a stratum or population have `something
in common', modelled as $\corr(X_i,X_j)=\rho$ for all $i\not=j$.
In that case,
\beq
\E\,M_n\doteq\xi+(1-\rho)^{1/2}\sigma(2\log n)^{1/2}, 
\label{eq:maximummeannormal}
\eeq 
so with more similarity than under pure independence,
the maxima follow the same pattern, modulo a reduction factor;
more independence between individuals imply bigger extremes. 

For illustration, consider countries A and B where something
of sufficient interest is recorded for their people 
have distributions $\N(\xi,\sigma_A^2)$ and $\N(\xi,\sigma_B^2)$.
The `winners' from A and B can be expected to have scores
\beqn
\xi+\sigma_A(2\log n_A)^{1/2}
\quadandquad
\xi+\sigma_B(2\log n_B)^{1/2}, 
\eeqn 
in terms of population sizes $n_A$ and $n_B$. The excess ratio,
B compared to A, is essentially 
\beqn
r={\sigma_B(\log n_B)^{1/2} \over \sigma_A(\log n_A)^{1/2}}. 
\eeqn
For the participation ratio, note that if A is e.g.~3 times
bigger than B, the sample size ratio is
\beqn
    {(\log 3+\log n_B)^{1/2}\over (\log n_B)^{1/2}}
    =\Bigl(1+{\log 3\over \log n_B}\Bigr)^{1/2}, 
\eeqn 
indicating that the difference matters, when it comes
to the highest-scoring levels, but not by much,
when it comes to very high sizes. The tallest Dutch
cannot be expected to be much taller than the tallest Norwegians,
assuming here these countries have the same grand average
and the same grand variances, even if the Duch population
s three times that of the Norwegians.
Neither is it entirely impossible that a guy from Lommedalen
is taller than all Russians. 

In many cases, therefore, the standard deviation ratio
matters rather more than the sample sizes. 
To get even with B, A needs 
\beqn
    {\log n_B\over \log n_A} = {\sigma_A^2\over \sigma_B^2},
    \quad {\rm or\ }
    n_A=n_B\exp(\sigma_B^2/\sigma_A^2), 
\eeqn
featuring the crucial variance ratio. 
For illustration, suppose A and B have $\N(100,15.0^2)$ and $\N(100,16.0^2)$;
then B will indeed tend to have the tallest scores. 
Country A needs to be 15 percent bigger than B in order
to make up for the small 6 percent standard deviation increase. 

\section{A parametric model for ratings over threshold}
\label{section:paramodel} 

My two-parameter model for scores above threshold
has density of the form 
\beq
f(x,a,\theta)= \Gamma(a+1)^{-1}
\exp\Bigl\{ -\Bigl({x-r_0\over \theta}\Bigr)^{1/a} \Bigr\}{1\over \theta}, 
   \quad {\rm for\ } x\ge r_0, 
\label{eq:hereisf}
\eeq
monotonically decreasing on its domain. 
We shall see in Section \ref{section:chessratingjan2026}
that the fit to chess scores data is very good.
Here $a$ is driving the fine aspects of the upper tail,
whereas $\theta$ is a scale parameter reflecting the
variability of the distribution.

\subsection{Properties.} 

A little exercise verifies that the integration constant
in (\ref{eq:hereisf}) is the right one, and that the model also
can also be represented as
\beq
X=r_0+\theta\,V_a^a
\quad {\rm where\ }V_a\sim\Gam(a,1). 
\label{eq:hereisX} 
\eeq 
Formulae for moments and quantiles may be deduced from this.
In particular,
\beq
\E\,X=r_0+\theta{\Gamma(2a)\over \Gamma(a)},
\quad
\Var\,X=\theta^2 \Bigl[{\Gamma(3a)\over \Gamma(a)}
     -\Bigl\{{\Gamma(2a)\over \Gamma(a)}\Bigr\}^2\Bigr]. 
\label{eq:meanvariance} 
\eeq 
Furthermore, solving $F(x,a,\theta)=p$ for $x$ leads to
the quantile function
\beq
F^{-1}(p,a,\theta)=r_0+\theta\,G_a^{-1}(p)^a. 
\label{eq:quantilefunction}
\eeq 
Also useful is an explicit formula for the cumulative
distribution function (c.d.f.), via (\ref{eq:hereisX}):
\beq
F(x,a,\theta)=\Pr(r_0+\theta\,V_a^a\le x)
=G_a(((x-r_0)/\theta)^{1/a}) \quad {\rm for\ }x\ge r_0, 
\label{eq:hereisF}
\eeq 
in terms of the c.d.f.~$G_a$ for the $\Gam(a,1)$. 

The maxima of samples of players from some given stratum,
like all active players in a country, follow the pattern
of (\ref{eq:maximummeannormal}),  
though with an adjusted logarithmic speed, depending
on the $a$. Specifically, with $M_n$ the maximum
of an i.i.d.~sample from (\ref{eq:hereisf}), we have 
\beq
\E\,M_n\doteq r_0+\theta (\log n)^a 
\label{eq:maximamean} 
\eeq 
for growing participation volume $n$. The Gaussian case
corresponds to $a=\half$ and the exponential to $a=1$;
the chess scores have tails between these two classical
cases. Formula (\ref{eq:maximamean}) follows from
extreme value theory; we may point to 
\citet{Gumbel19,Gumbel22,Gumbel58},
\citet{Embrechtsetal97}. There is a Gumbel distribution
limit for the maximum here, with
$(M_n-a_n)/b_n\arr_d G$, the Gumbel distribution
with c.d.f.~$\exp(-\exp(-g))$, for suitable slowly
increasing $a_n$ and $b_n$, with consequences
also for all quantiles; see the details in
Remark B of Section \ref{section:concluding}. 

\subsection{Estimation and inference.} 

Consider a stratum of interest, with ratings $x_1,\ldots,x_n$
for its $n$ individuals. We need to estimate the parameters
$(a,\theta)$ in the (\ref{eq:hereisf}) model.  

Suppose first that the full set $x_1,\ldots,x_n$ is available.
The log-likelihood is then
\beq
\ell(a,\theta)=\sumin \Bigl\{-\log\Gamma(a+1)
-\Bigr({x_i-r_0\over \theta}\Bigr)^{1/a}-\log\theta\Bigr\},
\label{eq:logL}
\eeq 
which can be maximised numerically. When needed, for numerical
precision or for bigger datasets, the maximisation can be
reduced to a function of $a$ alone, namely $\ell(a,\hatt\theta(a))$, 
as the log-likelihood for given $a$ is maximised for
\beqn
\hatt\theta(a)=n^{-1}\sumin (x_i-r_0)^{1/a}. 
\eeqn 

Intriguingly and importantly, parameters may be estimated also in cases
where only the say the top-100 ratings are listed; in various
databases for chess and performance sports only such top-$k$ lists
are available, for suitable cutoff $k$. Sorting these top $k$
values as $x_1>\cdots>x_k$, then, one merely knows that
the other $n-k$ ratings are inside the $[r_0,x_k]$ box.
The statistical information, clearly less complete compared
to knowing all the data, can be represented by the likelihood
\beqn
L_k(a,\theta)=F(x_k,a,\theta)^{n-k} f(x_k,\theta)\cdots f(x_1,\theta), 
\eeqn 
i.e.~with log-likelihood
\beq
\ell_k(a,\theta)=(n-k)\log F(x_k,a,\theta)^{n-k}
   +\sum_{i=1}^k \log f(x_k,a,\theta), 
\label{eq:loglikelihoodB}
\eeq
which again can be maximised to fit the available top-$k$ dataset.
This takes knowing the participation volume, i.e.~the $n$, though. 

\section{Chess ratings, for men and women,
  per January 2026} 
\label{section:chessratingjan2026}

From the FIDE website I have got hold of the 15,424
players, active as of January 2026, with regular ratings
equal to or above the threshold $r_0=2100$,
with $n_m=14671$ male and $n_w=753$ female players.
Top-rated Magnus Carlsen has 2840, whereas
Hou Yifan with 2613 leads the women table;
we note, for further analysis below, that the top-100 rated
players are all men. 

We can now fit the parameters of the (\ref{eq:hereisf})
density by maximising the log-likelihood (\ref{eq:logL}),
for the two categories. Approximate standard deviations
may also be computed, for the basic model parameters
and for smooth functions thereof, via the delta method;
see e.g.~\citet[Chs.~2, 5]{HjortStoltenberg26}.
In some detail, for the maximum likelihood (ML) estimators, 
\beqn
\begin{pmatrix}
  \hatt a \\ \hatt\theta 
\end{pmatrix}\approx_d\N_2(
\begin{pmatrix}
  a \\ \theta 
\end{pmatrix},\hatt J^{-1}), 
\eeqn 
with $\hatt J$ the observed Fisher information matrix,
i.e.~minus the Hessian at the ML position. Also, for
any focus parameter $\mu=\mu(a,\theta)$, like the mean,
the standard deviation, or a quantile, we have 
\beqn
\hatt\mu=\mu(\hatt a,\hatt\theta)\approx_d\N(\mu,\hatt\tau^2),
\quad {\rm with\ }\hatt\tau^2=\hatt c^\tr \hatt J^{-1}\hatt c, 
\eeqn 
where $\hatt c$ is the gradient of the $\mu$ evaluated at the ML.
Results are as follows, for the two strata,
with estimates and standard errors (estimated standard deviations): 

\begin{small}
\begin{verbatim}
Table I: 
         men              women
a          0.689 0.013      0.612  0.052
theta    209.28  3.49     194.86  12.74
mu      2241.52  0.99    2221.53   3.60
sigma    119.67  1.07      98.31   1.58 
median  2210.79  1.05    2198.23   3.98 
\end{verbatim} 
\end{small}


\begin{figure}[h]
\centering
\includegraphics[scale=0.35]{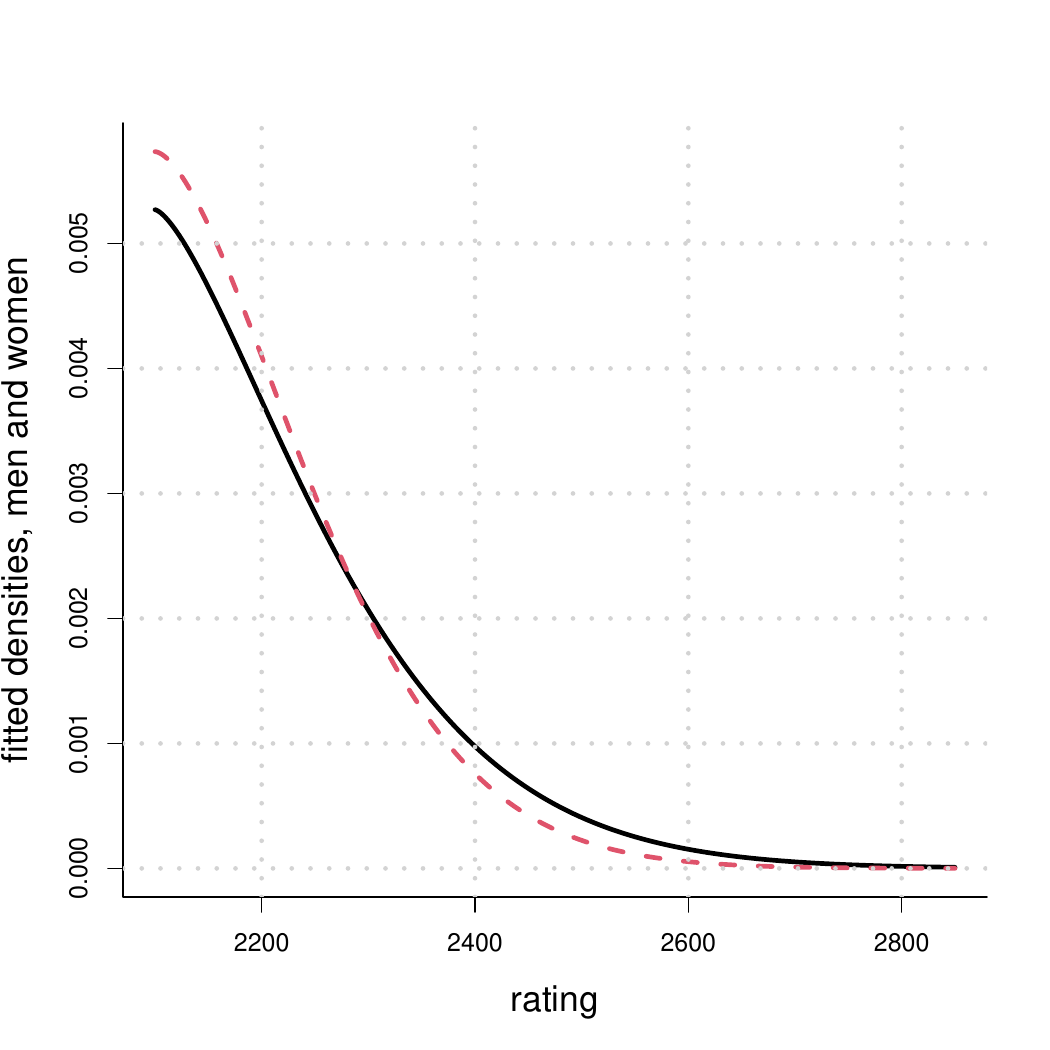}
\includegraphics[scale=0.35]{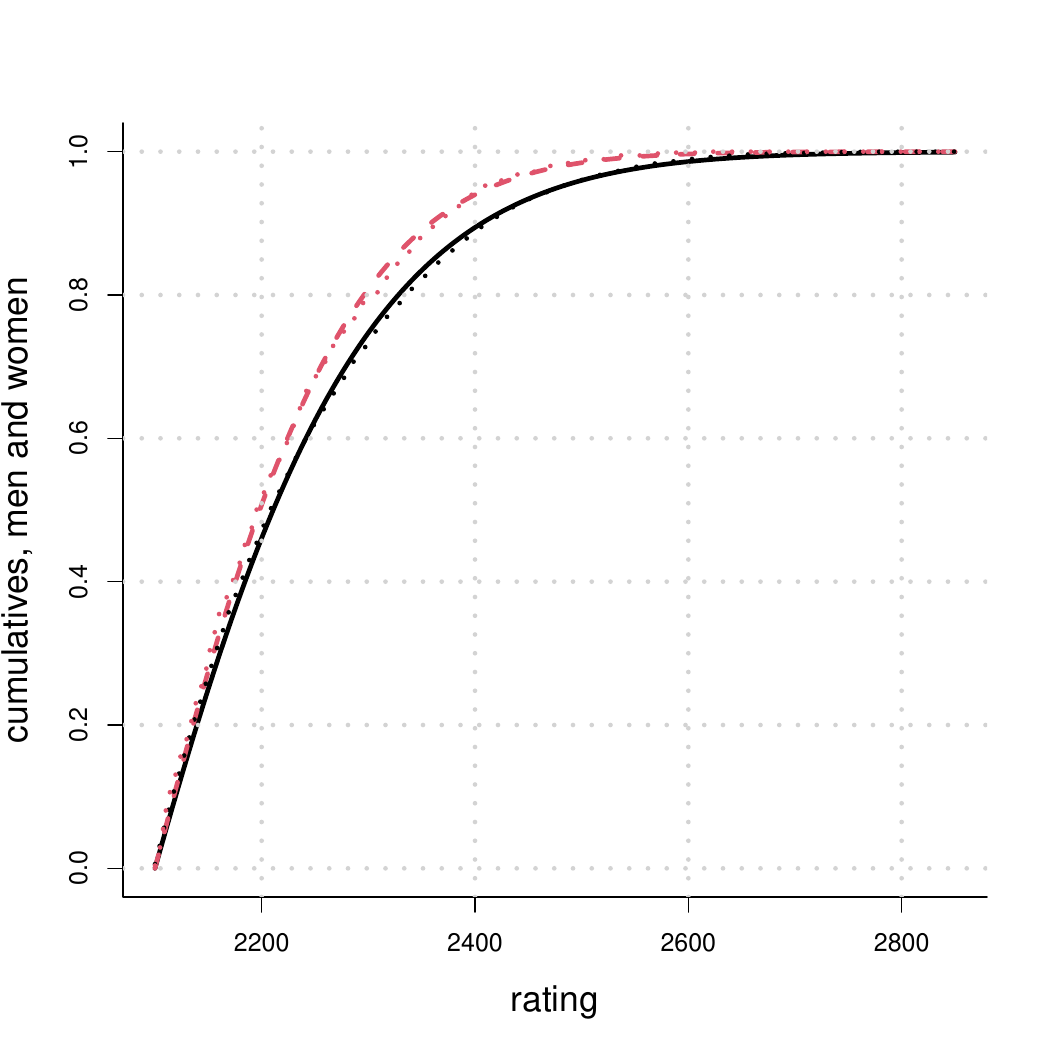}
\caption{
  Left panel: fitted densities $\hatt f_m$ and $\hatt f_w$,
  for men (full, black) and women (red, slanted)
  having rating above 2100;
  they are not far apart but the right tail is dominated
  by the men. 
  Right panel: fitted c.d.f.s $\hatt F_m$ and $\hatt F_w$,
  along also with the empirical c.d.f.s~(very slightly non-continuous,
  as dotted curves). The parametric and nonparametric
  are very close, i.e.~the model used is fully adequate.}
\label{figure:fig21}
\end{figure}

The estimated densities $\hatt f_m$ and $\hatt f_w$
for men and women are displayed in Figure \ref{figure:fig21},
left panel; these are seen to be close for the main bulk of
players, but with a significantly fatter right tail for the men,
That aspect, which shows up also for c.d.f.s and quantile
functions, is not coupled with the higher participation
number for men, per se, but by the upper performances
of the two groups. A mild caveat, though, is that the
estimated curves and quantities for the women group
are less precise than those for the men. 

In Figure \ref{figure:fig21}, right panel,
both the parametric and empirical c.d.f.s are portrayed, i.e.
\beqn
\hatt F_m(x)&=&F(x,\hatt a_m,\hatt\theta_m), \hspace{1.65cm} \quad 
\hatt F_w(x)=F(x,\hatt a_w,\hatt\theta_w), \\ 
F_{\emp,m}(x)&=&n_m^{-1}\sum_{i=1}^{n_m} I(x_{m,i}\le x), \quad
F_{\emp,w}(x)=n_w^{-1}\sum_{i=1}^{n_w} I(x_{w,i}\le x). 
\eeqn 
We learn from this that the model (\ref{eq:hereisf})
is even surprisingly adequate; formal monitoring of
$\hatt F_w - F_{\emp,w}$ shows no significant deviation from zero, etc.

\begin{figure}[h]
\centering
\includegraphics[scale=0.35]{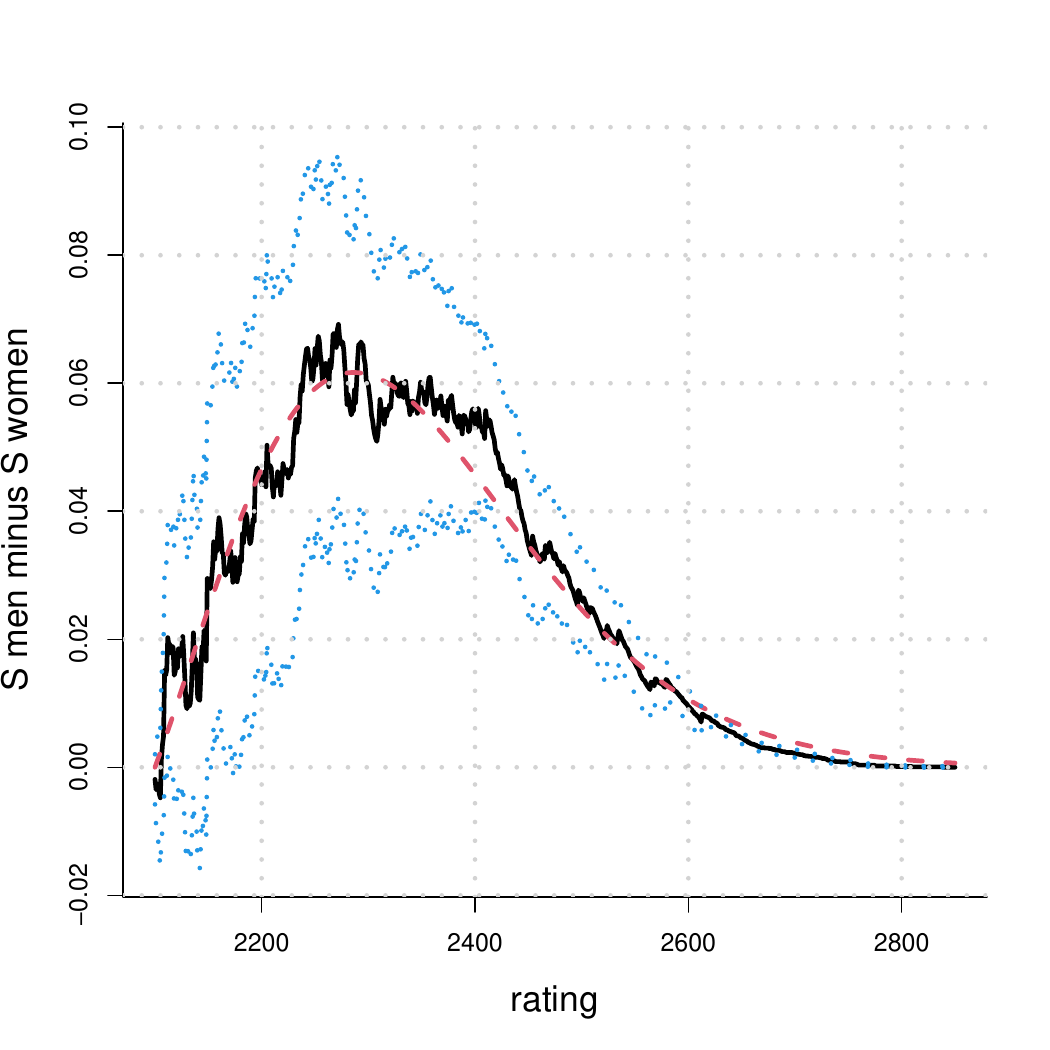}
\includegraphics[scale=0.35]{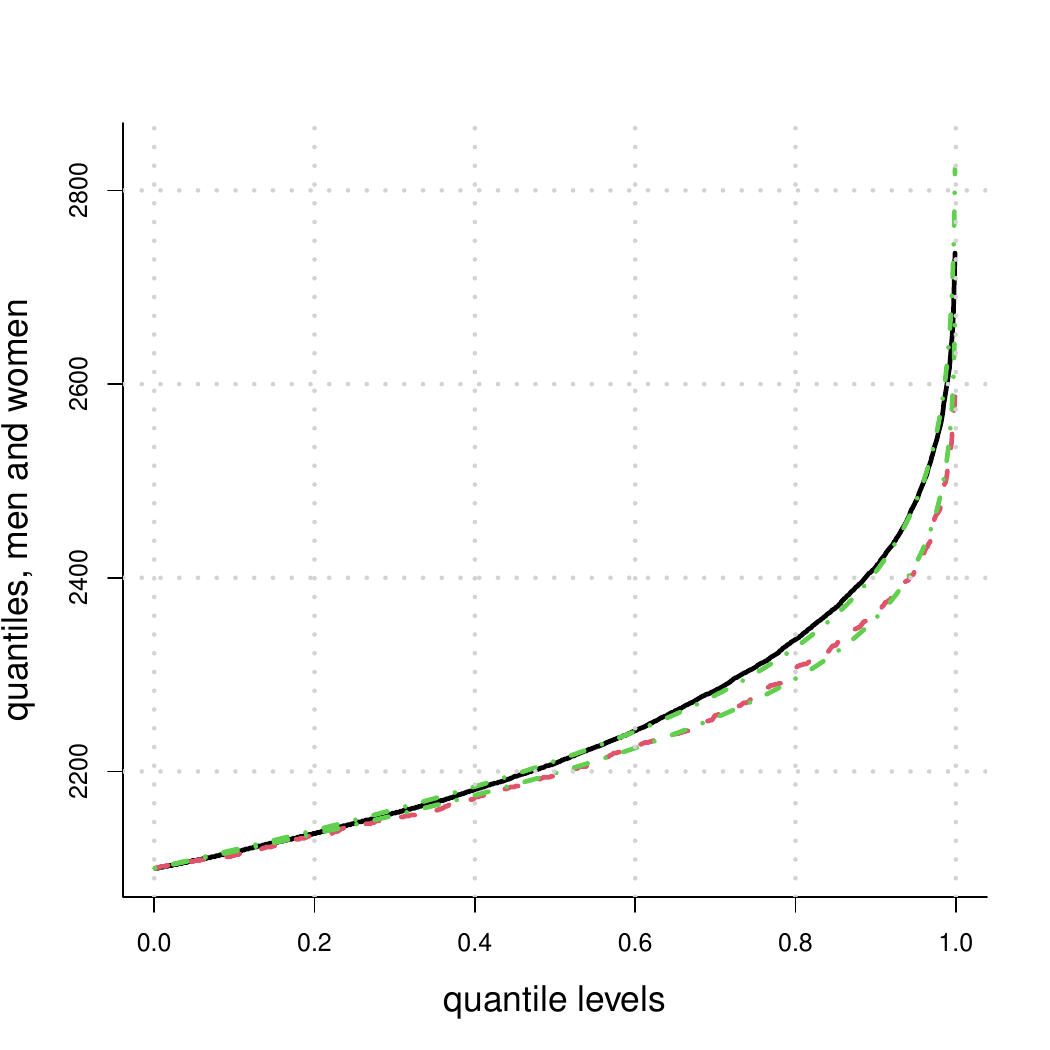}
\caption{
  Left panel: the difference $S_m(x)-S_w(x)$,
  empirical (black, wiggly) and parametrically fitted
  (red, slanted, smooth), along with a 90 percent
  confidence band.
  Right panel: empirical and theoretical quantile
  functions $Q_m$ (black, full) and $Q_w$ (red, slanted).
  The model fit is excellent.}
\label{figure:fig22}
\end{figure}

In Figure \ref{figure:fig22}, left panel, we also plot
the difference $S_{\emp,m} - S_{\emp,w}$ (black, wiggly), with
$S_{\emp,m}(x)$ and $S_{\emp,w}(x)$ defined as the the remaining
factions above $x$. So at all levels $x$ the male proportion
above it is a few percent more than for the femails.
The red slanted curve is the corresponding parametrically
estimated difference, again indicating a very good fit.
The lower and upper blue wiggly curves form a 90 percent
confidence band for the $S_m-S_w$ difference, indicating
also that the differences noted are statistically significant. 
Finally, that figure's right panel displays the estimated
quantile functions, again with the parametric fits very
close to the empirical ones. Visibly and significantly,
there is a gender gap at the very top levels
(for the populations studied, those above 2100 regular rating,
for those active as of January 2026). 

\section{Assessing the gender gap} 
\label{section:testing}



\begin{figure}[h]
\centering
\includegraphics[scale=0.35]{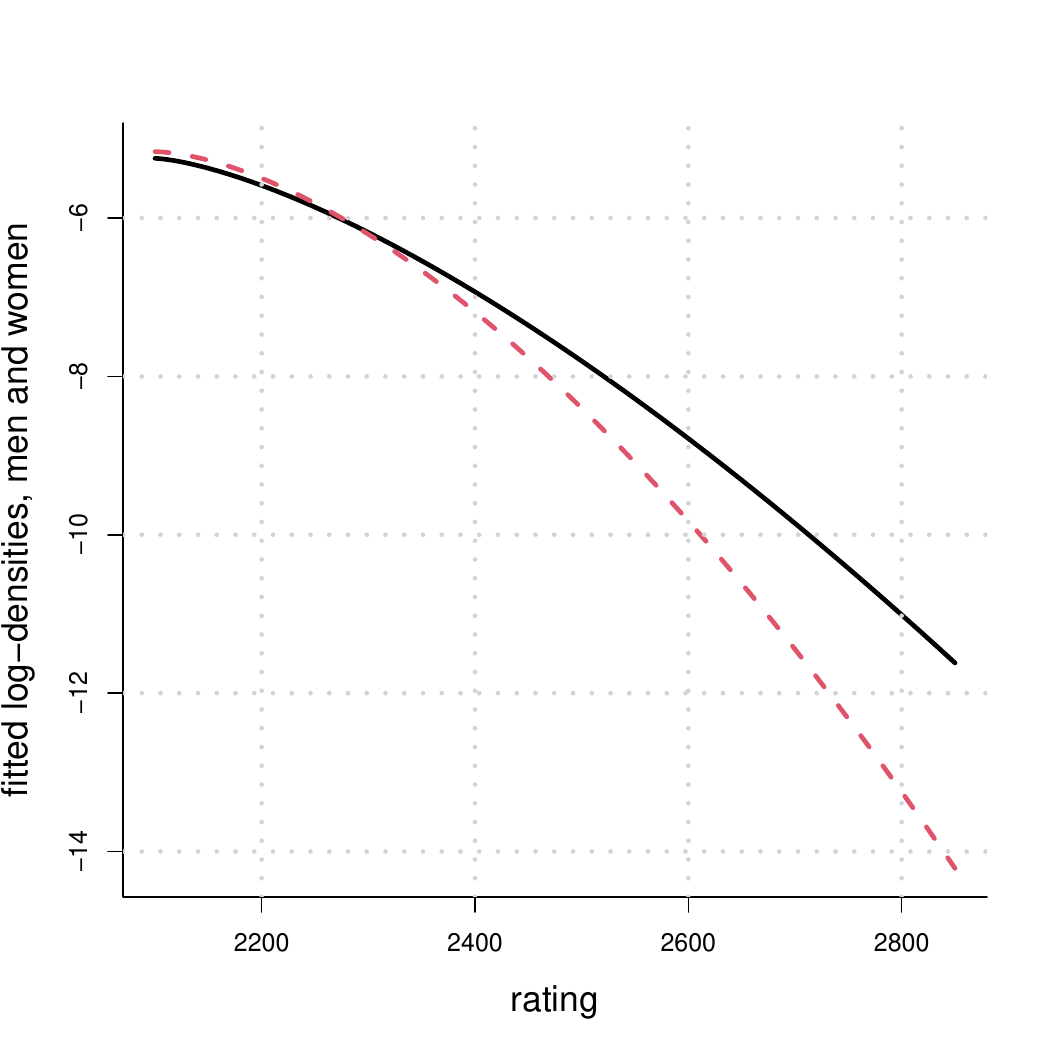}
\includegraphics[scale=0.35]{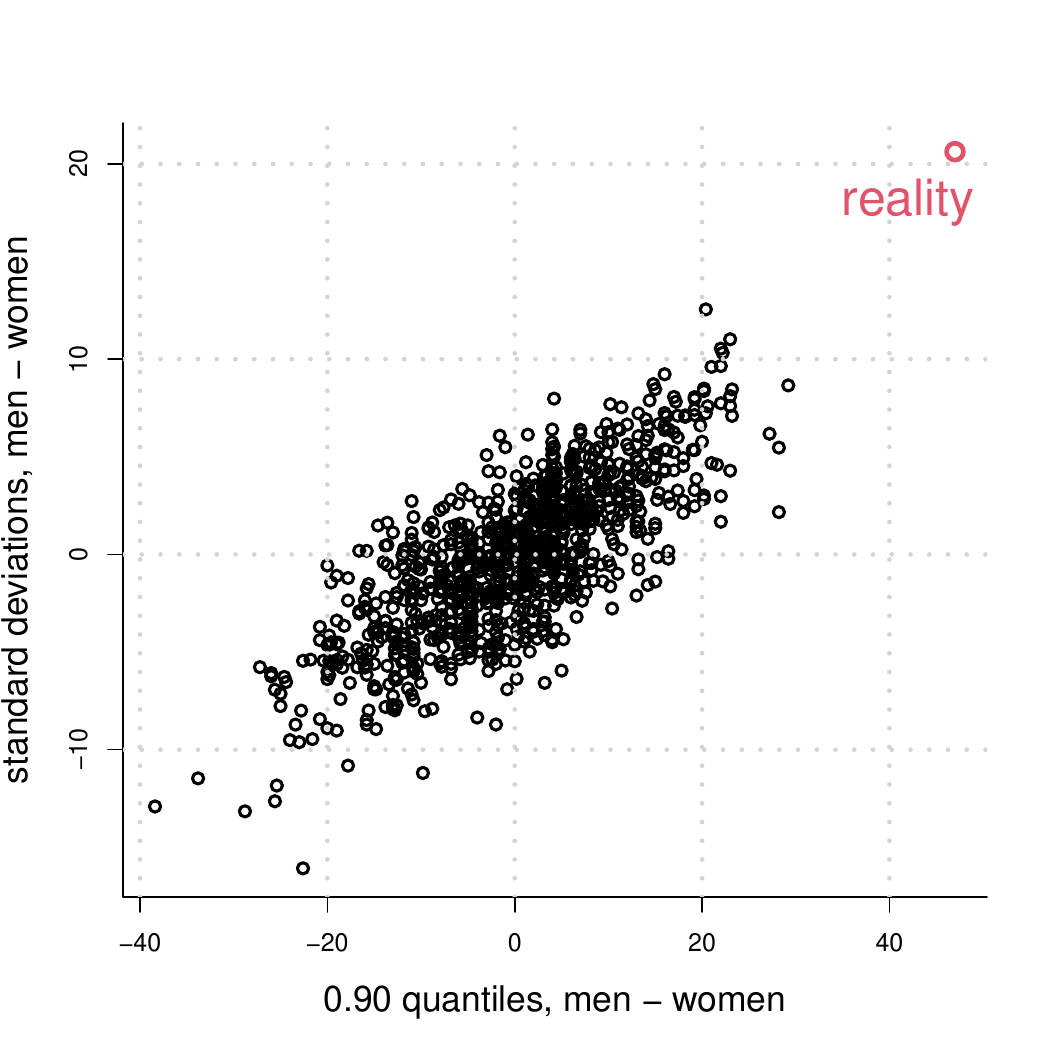}
\caption{
  Left panel: the estimated log-densities $\log\hatt f_m$
  and $\log\hatt f_w$, indicating that the man tail is
  fatter and longer than the woman tail; more top players
  are, indeed, men.
  Right panel: bootstrap points $(A^*,B^*)$ for 1,000
  bootstrap null distribution datasets, along with
  the factual $(A_\obs,B_\obs)$, for $A$ the 0.90 quantile
  difference and $B$ the standard deviation difference.}
\label{figure:fig26} 
\end{figure}

We have seen above, see e.g.~Table I, that the $a_m=a_w$
hypothesis is not in conflict with the data. 
For the two standard deviation estimates 119.67 and 98.31, 
however, with standard errors 1.07 and 1.48, we have
\beqn
t={\hatt\theta_m-\hatt\theta_w\over (\se_m^2+\se_w^2)^{1/2}}
   ={21.26\over 1.91}, 
\eeqn 
indicating that the male variance is much stronger than the female one.
This also agrees with our findings and figures above.
Whereas Figure \ref{figure:fig21}, left panel, shows
the estimated direct densities $\hatt f_m,\hatt f_w$,
Figure \ref{figure:fig26}, left panel, shows
their logarithms, i.e.~$\log\hatt f_m$ and $\log\hatt f_w$,
showing more clearly that the male tail is bigger and fatter
than the female tale. 

We hence already know well that there are significant
differences between the top men and top women players,
for regular chess ratings, as of January 2026.
It is instructive to build other formal tests as well.
A class of such tests take the following form,
using bootstrap samples.
Pooling the men and women together, with $n=n_m+n_w=15424$
chess scores, sample randomly $n_m=14671$ of these,
labelled $x_m^*$, the remaining $n_w=753$ labelled $x_w^*$.
From such a random bootstrap split of the full dataset,
we may compute discrepancy scores, like
\beqn
A=(F_{\emp,m}^*)^{-1}(0.90) - (F_{\emp,w}^*)^{-1}(0.90)
\quadandquad
B=\sd(x_m^*)-\sd(x_w^*),
\eeqn 
the differences of 0.90 quantiles and of standard deviations.
Doing this say 1,000 times yields bootstrap null distributions
of $A$ and $B$, under the no-difference hypothesis.
This is portrayed in Figure \ref{figure:fig26}, right panel. 
Then we compute the actually observed $(A_\obs,B_\obs)=(47.0,20.6)$,
plotted in the right up corner of that figure.
We see again, in a clear fashion, that the null hypothesis
of gender equality is firmly rejected;
the differences $A_\obs$ and $B_\obs$ are far outside
the null hypothesis range.

\section{Concluding remarks} 
\label{section:concluding}

In this note I have introduced a two-parameter model
for scores above thresholds, and demonstrated
that it very adequately describes the statistics
of top players in chess. The model, with relevantly
estimated parameters, may be used to monitor
progress over time, to predict overall level performances
in tournaments, and can surely be applied to other
top scores phenomena, in biology or otherwise. 

To round off this note a few comments are as follows. 

{\it Remark A. Same tail index for men and women.}
Slightly sharper estimates might be given,
than those obtained in Table I
of Section \ref{section:chessratingjan2026}, 
via estimation in the three-parameter
model which takes $a_m=a_w$; model selection analysis,
with the AIC and the BIC, cf.~\citet[Chs.~2,3]{ClaeskensHjort08},
indicate that this three-parameter reduction
has a slightly better explanatory power than
needing two plus two parameters. It is also of interest
to see in this fashion that the tail parameters are quite close
for the two groups (the null hypothesis $a_m=a_w$ is not
refuted), indicating that the primary factor
underlying the current gender gap in top chess
is not participation, or different skills for the
main bulk of top players, but variability.

{\it Remark B. Sample extremes.}
Suppose $u_1,\ldots,u_n$ are an i.i.d.~sample for
the density $f(u)=\Gamma(a+1)^{-1}\exp(-u^{1/a})$ on the
half{}line, with tail behaviour dictated
by the $a$ parameter. For large $u$, the survivor function is 
\beqn
\bar F(u)=1-F(u)\doteq cu^{1-1/a}\exp(-u^{1/a}), 
\eeqn
with the dominant term being the stretched exponential.
The tail is Gaussian for $a=\half$ and exponential for $a=1$. 
The sample winner $M_n=\maxin u_i$ will have
$\E\,M_n\doteq (\log n)^a$. 
More precisely, there is a Gumbel limit, 
\beqn
    (M_n-b_n)/a_n \arr_d G,\quad {\rm with\ }
    a_n=a(\log n)^{a-1},\,\, b_n=(\log n)^a, 
\eeqn 
which leads to
\beqn
\E\,M_n=(\log n)^a + \gamma_e a(\log n)^{a-1} + o((\log n)^{a-1}), 
\eeqn 
with $\gamma_e=0.5772... $ the Euler constant. 

For our chess ratings model, then, in view of the
representation (\ref{eq:hereisX}), the expected
top rating, for a stratum with $n$ players above 2100, is
\beqn
\E\,M_{n,m}&\doteq& 2210 + 210.22\,(\log n)^{0.68}, \\
\E\,M_{n,w}&\doteq& 2200 + 179.04\,(\log n)^{0.61},
\eeqn 
for men and women.
These estimates, or predictions for future years or other
strata, can be sharpened somewhat in view
of Remark A. 

{\it Remark C. The Rashomon Syndrome}.
Statistical modelling and inference can be very good
at explaining the statistical essence of what goes on
for some perhaps complicated phenomenon studied,
in terms of pointing to mechanisms, predicting
what happens next, spotting and assessing differences
between groups, etc. etc. A sound caveat is that
of the so-called Rashomon Syndrome, however; 
rather different statistical `explanations',
from perhaps different models and arguments,
might turn out to be explaining the data equally well. 
So even though the (\ref{eq:hereisf}) model succeeds
very well in fitting all such higher ratings chess
data, it should not tbe seen as the canonical
`explanation' for what goes on in the minds of the
15,000 most clever chess players of our time.  

{\it Remark D. Learning the full iceberg
  from only seeing the top.}
As explained with (\ref{eq:loglikelihoodB}),
the full model can be learned from recording
say only the top-100, as long as we also know
the participation level, i.e.~the size $n$ above
the threshold in question. This should be useful
when it comes to analysing chess ratings
for various strata or groups for which the full
datasets are not fully known. 

{\it Remark E. Other gender gap literature.}
There are many articles and stories, published
in journals or as blog posts or expert views
in social media, pertaining to the gender gap
in chess; the extent to which it is present,
what likely causes might be, and how to amend,
in the direction of less imbalance. A simple search
on Google Scholar finds more than a hundred publications.
Good articles published on the web include
{\it What gender gap in chess?}, by Wei Ji Ma,
who argues that participation is the dominating
factor. The present note does not disagree with this,
but points to the crucial presence of {\it statistical variation}
within strata. 

{\it Remark F. Application to other domains.}
Clearly, `high scores above threshold' extreme value
data abound in other application areas, from finance
and meteorology and sports,
see e.g.~\citet{Hjort26}, to biological traits,
like heights and academic skills.
Note e.g.~via Remark D that if we know the tallest
50 US basketball players, in a season with 500 players,
we may actually infer the full distribution. 



\section*{Acknowledgments} 

I greatly enjoyed following the Norwegian Broadcasting
Corporation's home channel televising the World Rapid and
World Blitz from Doha, Qatar, during the Christmas week of 2025,
captained by Torstein Bae, pedagogically guiding us through
the jungles of complexities towards
Magnus Carlsen's magnificent wins.
I am also indebted to Arne Tj\o lsen who volunteered
assisting me in getting hold of the right data-files from FIDE. 

\begin{small}

\bibliographystyle{biometrika}
\bibliography{diverse_bibliography2026}

\end{small}

\bigskip
\bigskip

\begin{footnotesize} 

\noindent 
{\it Footnote.} After having been absorbed by the most exciting chess
in `10,000 Lifetimes' during the Christmas Week of 2025
I was inspired to dig out data, construct and test my models
(along with a few discarded attempts, like the Pareto
and the truncated normal, neither of which work well here).
I then wished to write up a good {\it FocuStat Blog Post},
as these are sometimes reaching a sizeable international
readership (we've got fanmail from Steven Pinker, etc.).
To my frustration {\it\&} irritation the University of Oslo
has changed the Wondrously Well-Working Vortex {\tt html},
which I and FocuStat comrades have used for ten years,
for a Horribly Bad-Working Vortex editing system, where
writing math is a nightmare; having the new Vortex system
installed has also seriously disturbed some of our previously
perfect pitch layouted Blog Posts. I might have to move
to Andorra to find a university taking better care
of its humble employees. 

\end{footnotesize} 

\end{document}